\documentclass[amsfonts,amsmath,prd,preprint,nofootinbib]{revtex4}
\newcommand{\beq}{\begin{equation}}
\newcommand{\eeq}{\end{equation}}

\newcommand{\bsp}{\begin{split}}

\usepackage{epsfig,bbm,cancel,ulem}
\usepackage[breaklinks=true]{hyperref}
\usepackage{xcolor}
\usepackage{wasysym}
\usepackage[mathscr]{eucal}
\usepackage{multirow}
\usepackage{amsmath}
\usepackage{enumitem}

\usepackage{slashed}

\begin{document}

\title{Hidden BPS states of  electroweak monopole and a new bound estimate}

\author{A.~Gunawan}
\email{alfian.gunawan@sci.ui.ac.id}
\author{H.~S.~Ramadhan\footnote{Corresponding author.}}
\email{hramad@sci.ui.ac.id}
\affiliation{Departemen Fisika, FMIPA, Universitas Indonesia, Depok, 16424, Indonesia. }

\author{I.~Prasetyo}
\email{ilham.prasetyo@sampoernauniversity.ac.id}
\affiliation{ Department of General Education, Faculty of Art and Sciences, Sampoerna University, Jakarta, 12780, Indonesia. }

\def\changenote#1{\footnote{\bf #1}}

\begin{abstract}

Using the BPS Lagrangian method, we obtain a distinct set of Bogomolny equations for the Cho-Maison monopoles from the bosonic sector of a regularized electroweak theory. In the limit of 
$n\rightarrow\infty$ of the permittivity regulator, $\epsilon\left(\rho^n\right)$, the mass of the monopole can be estimated to be $M_W\sim3.56$ TeV. This value is within the latest theoretical window, 2.98 TeV - 3.75 TeV. We also discuss some possible regularization mechanisms of electroweak monopole in the Yang-Mills sector and the existence of its BPS state.

\end{abstract}

\maketitle
\thispagestyle{empty}
\setcounter{page}{1}

\section{Introduction}
One of the intriguing asymmetry in nature is the empirical absence of magnetic monopole. Its non-existence breaks the Maxwellian duality symmetry. Their theoretical existence {\it can} nevertheless be put by hand in the Abelian theory. It was Dirac~\cite{Dirac:1948um} who first seriously considered the monopole's physical existence by showing that it can explain how electric charge is quantized. In doing so, he necessarily introduced the existence of an unphysical string describing the singularity of the monopole. This singularity seems to be an artifact of the Abelian gauge group.

While the monopole {\it can} be added in the Abelian gauge theory, they {\it must} exist in certain non-Abelian gauge formalisms. Wu and Yang~\cite{Wu:1976ge} proposed magnetic monopole without singular string using the non-aAbelian $SU(2)$ gauge theory. Later, 't Hooft \cite{tHooft:1974kcl} and Polyakov \cite{Polyakov:1974ek} introduced magnetic monopole as topological soliton, which shows that the $SU(2)$ gauge theory allows finite energy monopole solution. For excellent review on magnetic monopoles as topological solitons see, for example,~\cite{Preskill:1984gd, Shnir:2005vvi}. Unfortunately the correct non-Abelian construction for electroweak unification in nature is based on the $SU(2)\times U(1)_Y$ Weinberg-Salam model~\cite{Weinberg:1967tq, Salam:1968rm} whose vacuum manifold is a $3$-sphere, $\mathcal{M}=SU(2)\times U(1)/U(1)\cong\mathbb{S}^3$. Thus, the second homotopy group is trivial, $\pi_2\left(\mathbb{S}^3\right)=\mathbb{I}$.

Upon closer inspection, it is revealed that the vacuum manifold of the bosonic sector of Weinberg-Salam theory can have a non-contractible loop~\cite{Manton:1983nd}, suggesting the existence of monopole defects. This is made precise by Cho and Maison~\cite{Cho:1996qd} who realized that the extra $U(1)$ hypercharge can be viewed as a gauged $C\mathbb{P}^1$, thus resulting in $\pi_2\left(C\mathbb{P}^1\right)=\mathbb{Z}$. The Cho-Maison (CM) monopole, however, suffers from the singularity of energy due to its Coulombian (point-like) $U(1)$ hypercharge. This singularity can be understood if we view the CM monopole as a hybrid between the $U(1)$ (singular) Dirac and the $SU(2)$ (regular) t'Hooft-Polyakov monopoles~\cite{Cho:2012bq, Cho:2013vba}. Inspite of it, recent investigation has shown that the perturbation spectra of this electroweak monopole do not contain any negative mode; thus stable~\cite{Gervalle:2022npx}. It was originally hoped that the infinite energy can be made finite by either: embedding it into a larger (say, $SU(5)$) GUT group~\cite{Dokos:1979vu}, or coupling it with gravity~\cite{Bais:1975gu}.

The experimental search for magnetic monopole has been run, for example by MoEDAL collaboration, extensively~\cite{Pinfold:1999sp,MoEDAL:2014ttp,  Mitsou:2015yaj, Katre:2016sjg, MoEDAL:2016lxh, MoEDAL:2019ort, MoEDAL:2021vix}. CM monopole becomes one of the prominent candidate since it emerges from the realistic electroweak unification group. However, to detect its existence we need its mass estimate, and it is not so trivial given the singularity of the energy. In~\cite{Cho:2013vba} Cho, Kim, and Yoon (CKY) proposed three ways to estimate the mass bound: {\it dimensional argument}, {\it scaling argument}, both of which are backed up by {\it ultra-violet (UV)  regularization}. Their calculation shows that the monopole mass is around $4$ TeV $\lesssim$ $M_W\lesssim7$ TeV, which should be accessible to the MoEDAL. The CKY UV-regularization is based on the non-minimal coupling between the Higgs and the $U(1)$-gauge fields, $\mathcal{L}\ni\epsilon\left(|H|\right)B_{\mu\nu}B^{\mu\nu}$ with $\varepsilon\left(|H|\right)\propto|H|^8$. Ellis, Mavromatos, and You (EMY)~\cite{Ellis:2016glu} found that such power-law $\epsilon$ {\it regulator} phenomenologically contradicts the Higgs decay $H\rightarrow\gamma\gamma$~\cite{TheATLASandCMSCollaborations:2015bln}. Instead, they proposed more general polynomial functions whose coefficients are determined by requiring that $\epsilon$ satisfies the Principle of Maximum Entropy (PME)~\cite{Ellis:2016glu}. This approach gives a theoretical prediction that the electroweak magnetic monopole can be observed with a mass $\lesssim5.5$ TeV.

It is well-known that in the limit of vanishing potential the monopole is stable with globally-minimum energy. This is the Bogomolny-Prasad-Sommerfield (BPS) limit~\cite{Bogomolny:1975de, Prasad:1975kr}. Blaschke and Bene\v{s} were the first to show that the CKY model has BPS state in its spectrum~\cite{Blaschke:2017pym}. Using the Bogomolny formalism~\cite{Bogomolny:1975de} they calculated the lower bound of the energy to be $M_W\sim2.37$ TeV. Employing different regularization mechanism, Zhang, Zhou, and Cho (ZZC) obtained an energy bound around 3.75 TeV~\cite{Zhang:2020xsg}, but in the BPS state it approaches 2.37 TeV. Interestingly, the ZZC model allows the existence of non-unique BPS states. With different Bogomolny equations they show that they could obtain the energy bound 2.98 TeV. The non-uniqueness of BPS state in ZZC model of electroweak monopole is truly remarkable. The t'Hooft-Polyakov monopole and all its non-canonical variants (see~\cite{Casana:2012un, Casana:2013lna, Ramadhan:2015qku}) have a single BPS eenergy lower bound. To the best of our knowledge, the CM electroweak monopole in this ZZC regularization model is the first solitons with non-unique BPS energy. Investigating further the other possible sector of BPS states in this ZZC model is preceisely the aim of this work.

In this paper we try to construct other possible BPS configurations in the ZZC electroweak monopole by employing one of the Bogomolny formalism. This paper is organized as follows. In Sect.~\ref{sec:review} we briefly review the electroweak monopole. To obtain the BPS equations we employ one of the Bogomolny formalism. This method is reviewed in Sect.~\ref{sec:method}. We apply the Bogomolny method to the electroweak monopole with electromagnetic permittivity regulator in Sect.~\ref{sec:permitt}. In Sect.~\ref{sec:BI} we discuss several alternative proposals for electroweak monopole regularization and the existence of their BPS states. Finally, in Sect.~\ref{sec:conc} we disucss our results.

\section{Electroweak Monopole: A Review}
\label{sec:review}

Here we briefly review the existence of magnetic monopole in electroweak theory. The discussion heavily follows Refs.~\cite{Cho:1996qd, Mavromatos:2020gwk}.

The electroweak theory is described by the 
$SU(2)\times U(1)_Y$ Weinberg-Salam model~\cite{Weinberg:1967tq, Salam:1968rm} whose Lagrangian consists of the following four parts:
\begin{equation}\label{lagrew}
\mathcal{L}_{EW}=\mathcal{L}_{h}+\mathcal{L}_{g}+\mathcal{L}_{f}+\mathcal{L}_{y}.
\end{equation}
Here $\mathcal{L}_{h}$, $\mathcal{L}_{g}$, $\mathcal{L}_{f}$, and $\mathcal{L}_{y}$ denote the scalar Englert-Brout-Higgs, the Abelian and non-Abelian gauge, the fermionic, and the interactions between the fermions Lagrangians, respectively,
\begin{eqnarray} 
    \mathcal{L}_{h}&=&-|\hat{D}_{\mu}H|^2-\frac{\lambda}{2}\left(H^\dagger H -\frac{\mu^2}{\lambda}\right)^2,\nonumber\\
    \mathcal{L}_{g}&=&-\frac{1}{4}F_{\mu\nu}^a F^{a\mu\nu}-\frac{1}{4} G_{\mu \nu}G^{\mu\nu},\nonumber\\
\mathcal{L}_{f}&=&\Bar{Q}_j i \slashed{D}Q_j+\bar{u}_j i \slashed{D}u_j +\bar{d}_j i \slashed{D} d_j +\bar{L}_j i \slashed{D}L_j +\bar{e}_j i \slashed{D} e_j,\nonumber\\
    \mathcal{L}_{y}&=&-y_{uij}\epsilon^{ab}h_b^\dagger \bar{Q}_{ia}u_j^c-y_{dij}h \bar{Q}_{i}d_j^c-y_{eij}h \bar{L}_{i}e_j^c + h.c.,\nonumber
\end{eqnarray}
where $\hat{D}_{\mu}$ is the $SU(2)\times U(1)_Y$ covariant derivative,
\begin{align*}
    \hat{D}_{\mu} H&\equiv \left(\partial_{\mu} - i\frac{g}{2}\sigma^a A_{\mu}^a-i\frac{g'}{2}B_{\mu}\right)H=\left(D_{\mu}-i\frac{g'}{2}B_{\mu}\right)H.
\end{align*}
The Englert-Brout-Higgs field $H$ is responsible for the $SU(2)\times U_Y (1)\rightarrow U_{\textrm{em}}(1)$ symmetry breaking, with $U_{\textrm{em}}(1)$ is the abelian gauge group for electromagnetism. The vector potential $A_\mu^a$ and $B_\mu$ in the field strength tensors $F_{\mu \nu}^a$ ($a=1,2,3$) and $G_{\mu \nu}$, respectively, are the gauge fields from $SU(2)$ and \textit{hypercharge} $U_Y(1)$.

The Cho-Maison monopole can be found in the bosonic sector of the Lagrangian~\eqref{lagrew}
\begin{eqnarray}
\label{monopolelagr1}
\mathcal{L}&=&\mathcal{L}_{h}+\mathcal{L}_{g}\nonumber\\
&=&-|\hat{D}_{\mu}H|^2-\frac{\lambda}{2}\left(H^\dagger H -\frac{\mu^2}{\lambda}\right)^2-\frac{1}{4}F_{\mu\nu}^a F^{a\mu\nu}-\frac{1}{4}G_{\mu\nu} G^{\mu\nu}\nonumber\\
&=&-\frac{1}{2}(\partial_{\mu} \rho)^2-\frac{\rho^2}{2}|D_{\mu} \xi|^2-\frac{\lambda}{8}(\rho^2-\rho_0^2)^2-\frac{1}{4}F_{\mu\nu}^a F^{a\mu\nu}-\frac{1}{4}G_{\mu\nu} G^{\mu\nu},
\end{eqnarray}
where in the last line the field $H$ is decomposed into the Higgs field $\rho$ and the unit doublet $\xi$, $H\equiv\left(\rho/\sqrt{2}\right)\xi$ with $\xi^\dagger \xi=1$. We also define $\rho_0\equiv \sqrt{2\mu^2/\lambda}$ to be the expectation value of the Higgs field in vacuum. The most general ansatz for CM monopole compatible with spherical symmetry is as follows~\cite{Cho:1996qd, Zhang:2020xsg, Mavromatos:2020gwk}:
\begin{eqnarray}
\label{ewansatz}
\rho&=&\rho(r),\quad\ \ \  \ \  \xi= i \begin{pmatrix}			\sin\left(\theta/2\right)e^{-i\varphi}\\
-\cos\left(\theta/2\right)
\end{pmatrix},\nonumber\\
\hat{n}&=&-\xi^\dagger \boldsymbol{\sigma}\xi=\hat{r},\quad C_{\mu}=-\frac{1}{g}\left(1-\cos\theta\right)\partial_\mu \varphi\nonumber\\
\mathbf{A}_{\mu}&=&\frac{1}{g}A(r)\partial_\mu t \hat{r}+\frac{1}{g}\left(f(r)-1\right)\hat{r}\times \partial_\mu \hat{r},\nonumber\\
B_{\mu}&=&\frac{1}{g'}B(r)\partial_\mu t -\frac{1}{g'}\left(1-\cos\theta\right)\partial_\mu \varphi.
\end{eqnarray}
Note that for the case of pure magnetic monopole, $A(r)=B(r)=0$. The gauge ansatz $A_\mu^a$ and $B_\mu$ in \eqref{ewansatz} can be written in the complete form:
\begin{eqnarray}
\mathbf{A}_t &=&(0,0,0) = \mathbf{A}_r, \nonumber\\
\mathbf{A}_{\theta}&=&\frac{1-f(r)}{g}\left(\sin\varphi,-\cos\varphi,0\right),\nonumber\\
\mathbf{A}_{\varphi}&=&\frac{1-f(r)}{g}\sin\theta \left(\cos\varphi\cos\theta,\sin\varphi \cos\theta,-\sin\theta\right),\nonumber\\
B_t&=&0=B_r=B_{\theta},\quad B_{\varphi}=-\frac{1}{g'}\left(1-\cos\theta\right).\nonumber
\end{eqnarray}

The unitary transformation $\xi\rightarrow U\xi=\begin{pmatrix}
		0\\1
\end{pmatrix}$ allows one to define physical gauge fields consisting of the electromagnetic field $A_\mu^{(\textrm{em})}$, the charged and the netral bosons $W_{\mu}$ $Z_\mu$, respectively:
\begin{eqnarray}
\label{ansatzem}
A_{\mu}^{(em)}&=&e\left(\frac{1}{g^2}A(r)+\frac{1}{{g'}^2}B(r)\right)\partial_{\mu} t-\frac{1}{e}\left(1-\cos\theta\right)\partial_{\mu} \varphi,\nonumber\\
W_{\mu}&=& \frac{i}{g}\frac{f(r)}{\sqrt{2}}e^{i\varphi} \left(\partial_{\mu} \theta + i\sin\theta_W \partial_{\mu} \varphi\right),\nonumber\\
Z_{\mu} &=&\frac{e}{gg'}\left(A(r)-B(r)\right)\partial_{\mu} t.
\end{eqnarray}
The Lagrangian \eqref{lagrew} in terms of the physical fields can be rewritten as
\begin{eqnarray} \label{transformedlagr}
\mathcal{L}&=&-\frac{1}{2}  \left(\partial_{\mu}\rho\right)^2-\frac{\lambda}{8}\left(\rho^2-\rho_0^2\right)^2-\frac{1}{4} {F'}^2_{\mu\nu}-\frac{1}{4}G_{\mu\nu}^2-\frac{1}{2}|D'_{\mu} W_{\nu} - D'_{\nu} W_{\mu}|^2\nonumber\\
&&-\frac{g^2}{4}\rho^2 W^*_{\mu} W^{\mu} - \frac{g^2+{g'}^2}{8}\rho^2 Z^2_{\mu}+ig F'_{\mu\nu} W^{*\mu}W^\nu+\frac{g^2}{4}\left(W^*_{\mu} W_{\nu}-W^*_{\nu} W_{\mu}\right)^2\nonumber\\
&=&-\frac{1}{2}\left(\partial_{\mu} \rho\right)^2-\frac{\lambda}{8}\left(\rho^2-\rho_0^2\right)^2-\frac{1}{4} {F_{\mu\nu}^{(\textrm{em})}}^2
-\frac{1}{4} Z_{\mu\nu}^2
-\frac{g^2}{4}\rho^2 W_{\mu}^* W^{\mu} -\frac{g^2+ {g'}^2}{8}\rho^2 Z_{\mu}^2 \nonumber\\
&&-\frac{1}{2}|\left(D_{\mu}^{(\textrm{em})}+i e\frac{g}{g'}Z_{\mu}\right)W_{\nu}-\left(D_{\nu}^{(\textrm{em})}+i e\frac{g}{g'}Z_{\nu}\right)W_{\mu}|
+i e \left(F_{\mu\nu}^{(\textrm{em})}+\frac{g}{g'}Z_{\mu\nu}\right){W^*}^{\mu} W^{\nu} \nonumber\\
&&+\frac{g^2}{4}\left(W_{\mu}^* W_{\nu}-W_{\nu}^* W_{\mu}\right),
\end{eqnarray} 
where 
\begin{eqnarray*}
D'_{\mu}&\equiv&\partial_{\mu} + i g A'_{\mu},\\ D_{\mu}^{\textrm{em}}&\equiv&\partial_{\mu} + i e A_{\mu}^{\textrm{(em)}}, \\
F'_{\mu \nu} &=& \partial_{\mu} A'_{\nu} - \partial_{\nu} A'_{\mu},\\
A'_{\mu} &=& A_{\mu} + C_{\mu},\\
C_{\mu} &=& -\frac{2 i}{g} \xi^\dagger \partial_{\mu} \xi,
\end{eqnarray*}
and $e$ (defined to be the electric charge) is related to the coupling constants $g$ and $g'$ by
\begin{equation}
e= \frac{g g'}{\sqrt{g^2 + {g'}^2}}.
\end{equation}
This corresponding field equations are
\begin{eqnarray} 
\label{eqn:2.5}
\rho''+\frac{2}{r}\rho'-\frac{1}{2r^2}f^2 \rho&=&\frac{\lambda}{2}\left(\rho^2-\rho_0^2\right)\rho -\frac{1}{4}\left(B-A\right)^2 \rho,\nonumber\\
f''-\frac{1}{r^2}\left(f^2-1\right)f&=&\frac{g^2}{4}\rho^2 f -A^2 f.
\end{eqnarray}
The simplest nontrivial solution of \eqref{eqn:2.5} is when $A(r)=B(r)=f(r)=0$, $\rho=\rho_0\equiv \sqrt{2} \mu/\sqrt{\lambda}$, and the EM gauge becomes
\begin{equation}
A_{\mu}^{(em)}=-\frac{1}{e}\left(1-\cos\theta\right) \partial_{\mu} \varphi,
\end{equation}
that is, a point-like magnetic monopole with the magnetic charge $4\pi/e$.

A more general solution can be obtained when $A(r)$, $B(r)$, and $f(r)$ are nonzero satisfying the boundary conditions
\begin{eqnarray}
\label{eq:bounddyon}
&\rho(0)=0,\qquad f(0)=1,\qquad A(0)=0, \qquad B(0)=b_0,\nonumber\\
&\rho(\infty)=\rho_0,\qquad f(\infty)=0,\qquad A(\infty)=B(\infty)=A_0,
\end{eqnarray}
where $0\leq A_0 \leq e\rho_0$ and $0\leq b_0 \leq A_0$. This condition describes a Cho-Maison dyon with the electric charge $q_e$ and magnetic charge $q_m$,
\begin{eqnarray}
q_e&=&-\frac{8\pi}{e}\sin^2\theta_W \int_0^\infty f^2 A\,dr=\frac{4\pi}{e}A_1, \nonumber\\
q_m&=&\frac{4\pi}{e}.
\end{eqnarray}
Here $A_1$ is the coefficient of $1/r$ term in the asymptotic expansion of $A(r)$,
\begin{equation}
A(r)\rightarrow A_0+\frac{A_1}{r}+\cdots.
\end{equation}

The ansatz \eqref{ansatzem} gives the energy of monopole  ($A(r)=B(r)=0$) in the form
\begin{eqnarray}
\label{eqenergy}
E&=&E_0+E_1,\\
E_0&\equiv& 4\pi \int_0^\infty \frac{dr}{2r^2}\left\{\frac{1}{g'^2}+ \frac{1}{g^2}\left(f^2-1\right)^2\right\},\\
E_1&\equiv& 4\pi \int_0^\infty dr \left\{\frac{1}{2}\left(r\rho'\right)^2+\frac{\lambda}{8}r^2 \left(\rho^2-\rho_0^2\right)^2+\frac{1}{g^2}f'^2+\frac{1}{4}f^2 \rho^2\right\}.
\end{eqnarray}
Applying the boundary conditions~\eqref{eq:bounddyon} for $\rho$ and $f$,
$E_1$ approaches a finite value
\begin{equation}
\label{E0}
E_1\approx 4.1 \textrm{ TeV},
\end{equation}
but $E_0\rightarrow\infty$ at the origin. This condition implies that there is no lower bound energy for this monopole, hence no BPS equations.

\section{BPS Lagrangian Method for Monopole}
\label{sec:method}

Bogomolny formalism is a convenient way of solving soliton's second-order field equations by reducing them into first-order. The corresponding equations and solutions are known as the BPS equations and BPS states, respectively. For monopole, the BPS state corresponds to the vanishing of the potential ($\lambda\rightarrow0$) while keeping the Higgs VEV finite, $\mu/\sqrt{\lambda}\rightarrow$ const.~\cite{Bogomolny:1975de, Prasad:1975kr}, thus preserving the non-trivial topology.  

The original Bogomolny mechanism was more of a {\it smart guess} trick of completing the square. Over the years there have been numerous proposals on constructing a systematic {\it algorithm} for obtaining the BPS equations (see, for example,~\cite{Sokalski:2001wk, Adam:2013hza, Atmaja:2014fha}). Among them is the so-called ``{\it BPS Lagrangian method}", initially used for some vortices models~\cite{Atmaja:2015umo}. This formalism has been used for obtaining BPS states of some exotic magnetic monopoles in~\cite{Atmaja:2018cod}. In this method, the BPS equations are the solutions of 
\begin{equation}
\label{l-l}
\mathcal{L}-\mathcal{L}_{BPS}=0.
\end{equation}
The brief review of this method is explained below.

In spherical coordinates where the system only depends on the radial coordinate $r$, the total static energy in the BPS limit is the difference of a BPS energy function in the limit 
$r\rightarrow\infty$ and $r\rightarrow 0$ \cite{Atmaja:2014fha}
	\begin{align}
		E_{BPS}=Q(r\rightarrow \infty)-Q(r\rightarrow 0)=\int_0^\infty dQ.
	\end{align}
According to the work by Atmaja, Prasetyo, and Ramadhan \cite{Atmaja:2015umo, Atmaja:2018cod}, $Q=\eta \pi P(\rho) F(f)$ is the suitable choice for BPS energy function in many models of vortices ($\eta=2$) and monopoles ($\eta=4$), where $\rho(r)$ and $f(r)$ are the components of the ansatz that depend on $r$. We can write $\mathcal{L}_{BPS}$ as
\begin{align}
\mathcal{L}_{BPS}&=-\frac{1}{4\pi r^2}\frac{dQ}{dr}=-\frac{1}{r^2}\frac{d}{dr}\left\{P(\rho)F(f)\right\}\nonumber\\
&=-\frac{1}{r^2}\left\{\frac{dP}{d\rho}\rho'(r) F(f)+P(\rho)\frac{d F}{d f}f'(r)\right\}\label{LBPS} ,
	\end{align}
with $\rho'(r)\equiv d\rho/dr$ and $f'(r)\equiv df/dr$. We can see that $\mathcal{L}_{BPS}$ consists of the first order derivative of $\rho(r)$ and $f(r)$, so Eq.~\eqref{l-l} is a polinomial of $\rho'(r)$ and $f'(r)$.

In the model where the static lagrangian of monopole is a quadratic function of $\rho'(r)$ and $f'(r)$, Eq.~\eqref{l-l} gives
\begin{eqnarray}
\label{blsexp}
\left\{\rho'(r)-G_1^{(1)}[\rho,f,f'(r)]\right\}\left\{\rho'(r)-G_1^{(2)}[\rho,f,f'(r)]\right\}&=&0 \quad \textrm{~or~} \nonumber\\
\left\{f'(r)-G_3^{(1)}[\rho,f,\rho'(r)]\right\}\left\{f'(r)-G_3^{(2)}[\rho,f,\rho'(r)]\right\}&=&0.
\end{eqnarray}
If we choose to solve roots of $\rho'(r)$ first, then to get a unique expression of $\rho'(r)$ we restrict $G_1^{(1)}-G_1^{(2)}=0$. This in turn can be rearranged to give us
\begin{equation}
\left\{f'(r)-G_2^{(1)}[\rho,f]\right\}\left\{f'(r)-G_2^{(2)}[\rho,f]\right\}=0,
\end{equation} 
which give us one expression of $f'(r)$ by $G_2^{(1)}-G_2^{(2)}=0$. This model must be valid for all $r$. Therefore, if $G_2^{(1)}-G_2^{(2)}$ is written as the polinomial of $r$,
\begin{equation}
G_2^{(1)}-G_2^{(2)}\equiv\sum_n a_n r^n=0,  
\end{equation}
we obtain some conditions for both $F(f) dP/d\rho$ and $P(\rho) dF/df$ from $a_n=0, \forall n$. The BPS equations are obtained from solving these conditions for both and substituting the results into $\rho'(r)$ and $f'(r)$.

Note that in general we cannot always express $\mathcal{L}_{BPS}$ proportionally as a total differential of the boundary term, as in Eq.~\eqref{LBPS}. A more general expression for $\mathcal{L}_{BPS}$ can be written as
\begin{equation}
\label{LBPS2}
\mathcal{L}_{BPS}=- Q_\rho(\rho,f) \frac{ \rho'(r)}{r^2}-Q_f(\rho,f) \frac{ f'(r)}{r^2},
\end{equation}
where $Q_\rho$ and $Q_f$ are in general functions of both $\rho$ and $f$. In this case, the total energy ($E$) and the topological charge ($Q$) cannot directly be inferred. The definition of topological charge, however, is still well-defined since it depends solely on the homotopy of the vacuum manifold. In this paper we shall encounter the BPS electroweak monopole where Eq.~\eqref{LBPS} does not hold.

\section{Bogomolny Formalism for Cho-Maison Monopole with Permittivity Regularization}
\label{sec:permitt}

As an attempt to regularize the singular CM monopole, ZZC~\cite{Zhang:2020xsg} proposed a non-canonical kinetic term on the $U(1)$ gauge sector of Lagrangian~\eqref{transformedlagr} by introducing a non-vacuum electromagnetic permittivity $\epsilon(\rho)$, 
\begin{eqnarray}
\label{eqn:2.11}
\mathcal{L}&=&-\frac{1}{2}\left(\partial_{\mu} \rho\right)^2-\frac{\lambda}{8}\left(\rho^2-\rho_0^2\right)^2-\frac{1}{4}\epsilon(\rho){F_{\mu\nu}^{(em)}}^2\nonumber\\
&&-\frac{1}{2}\left|\left(D_{\mu}^{(em)}+ie\frac{g}{g'}Z_{\mu}\right)W_{\nu}-\left(D_{\nu}^{(em)}+ie\frac{g}{g'}Z_{\nu}\right)W_{\mu} \right|^2\nonumber\\
&&-\frac{1}{4}Z_{\mu\nu}^2-\frac{g^2}{4}\rho^2 W_{\mu}^* W^{\mu} -\frac{g^2+g'^2}{8}\rho^2 Z_{\mu}^2\nonumber\\
&&+ie\left(F_{\mu \nu}^{(em)}+\frac{g}{g'}Z_{\mu \nu}\right){W^*}^{\mu} W^{\nu} +\frac{g^2}{4}\left(W^*_{\mu} W_{\nu}-W^*_{\nu} W_{\mu}\right)^2.
\end{eqnarray}
Using the ansatz~\eqref{ansatzem}, the monopole field equations are
\begin{eqnarray}
\label{eqn:2.12}
\rho''+\frac{2}{r}\rho'-\frac{1}{2r^2}f^2 \rho&=&\frac{\lambda}{2}\left(\rho^2-\rho_0^2\right)\rho -\frac{1}{4}\left(B-A\right)^2 \rho +\frac{\epsilon}{2}\left[\frac{1}{e^2 r^4}-e^2\left(\frac{A'}{g^2}+\frac{B'}{g'^2}\right)\right],
\nonumber\\ 
f''-\frac{1}{r^2}\left(f^2-1\right)f&=&\frac{g^2}{4}\rho^2 f -A^2 f.
\end{eqnarray}

The boundary condition is determined by considering the total energy:
\begin{eqnarray}
\label{eqn:2.14}
E&=&E_0+E_1,\\
E_0&\equiv& 4\pi \int_0^\infty \frac{dr}{2r^2}\left\{\frac{\epsilon}{e^2}+ \frac{1}{g^2}f^2\left(f^2-2\right)\right\},\\
E_1&\equiv& 4\pi \int_0^\infty dr \left\{\frac{1}{2}\left(r\rho'\right)^2+\frac{\lambda}{8}r^2\left(\rho^2-\rho_0^2\right)^2+\frac{1}{g^2}f'^2+\frac{1}{4}f^2 \rho^2\right\}.
\end{eqnarray}
It is evident that $E_0$ is divergent at the origin ($r\to 0$). This divergence can be removed by introducing a power-law function of $\epsilon(\rho)$  ~\cite{Zhang:2020xsg}:
\begin{eqnarray}
\epsilon&=&\left(\frac{\rho}{\rho_0}\right)^n\left\{c_0+c_1\left(\frac{\rho}{\rho_0}\right)+\cdots\right\},\\
\rho(r\to0)&\approx&r^\delta\left(h_0+h_1 r+\cdots\right),
\end{eqnarray}
where $\delta$ can be determined from Eq.~\eqref{eqn:2.12}:
\begin{align}
\delta=\frac{\sqrt{3}-1}{2},\qquad n>\frac{2}{\delta}=2\left(\sqrt{3}+1\right)\simeq 5.46.
\end{align}
In this condition, the first term in $E_0$ is finite near origin. To make the second term in $E_0$ finite, we need $f(0)=0$ or $\sqrt{2}$. However, both conditions do not result in finite energy. The solution to remove the singularity of Eq.~\eqref{eqn:2.14} is to combine both the divergent terms. Defining 
\begin{equation} \epsilon=\epsilon_0 +\epsilon_1,\qquad \epsilon_0=\frac{g'^2}{g^2+g'^2}=\frac{e^2}{g^2},\qquad \epsilon_1\simeq\left(\frac{\rho}{\rho_0}\right)^n,
\end{equation}
the singular energy $E_0$ can be written as
\begin{equation}
\label{limitzero}
E_0= 4\pi \int_0^\infty \frac{dr}{2e^2 r^2}\left\{\epsilon_0\left(f^2-1\right)^2+\epsilon_1\right\}.
\end{equation}
Thus, ZZC shows that the CM monopole energy is finite by taking $f(r\to0)=1$. This boundary condition is similar to the CKY case~\cite{Cho:2013vba}.
    
Applying the BPS Lagrangian method into this model, the Eq.~\eqref{limitzero} and the ansatz~\eqref{ansatzem}can be used to rewrite the Lagrangian as: 
\begin{equation}
\label{lagrs1}
\mathcal{L}=-\frac{{\rho'}^2}{2}- \frac{1}{2e^2 r^4}\left(\epsilon_0 \left(f^2-1\right)^2 +\epsilon_1 \right)-\frac{{f'}^2}{g^2 r^2}-\frac{f^2 \rho^2}{4r^2}-V(\rho).
\end{equation}
Here we add a potential term of the Higgs, $V(\rho)$. In  this method we do not a priori assume that $V=0$. Instead, such condition (to be precise, $V=$\ const.) appears as a constraint for the BPS equations to exist   (see, for example,~\cite{Atmaja:2015umo,Atmaja:2018cod}).


Putting the Lagrangians \eqref{LBPS2} and \eqref{lagrs1} into the Eq.~\eqref{l-l}, we end up with algebraic equations for $\rho'(r)$ and $f'(r)$. Solving $\rho'(r)$ first yields
\begin{equation}
\rho'_\pm=\frac{Q_\rho}{r^2}\pm \frac{1}{2r^2}\sqrt{-8r^4 V+4 Q_\rho^2-\frac{4}{e^2}\left(\epsilon_0\left(f^2-1\right)^2+\epsilon_1\right)-2r^2 f^2 \rho^2 +8 Q_f r^2 f'-\frac{8r^2 {f'}^2}{g^2}}.
\end{equation}
Demanding the two roots to be equal ($\rho'_+=\rho'_-$) results in a quadratic equation for $f'$, whose solution is
\begin{equation}
f'_\pm=\frac{g^2 Q_f}{2}\pm \frac{g}{er}\sqrt{e^2\left(2Q_\rho^2 +g^2 Q_f^2 r^2\right)-2\left(\epsilon_0 \left(f^2-1\right)^2+\epsilon_1\right)-e^2 r^2\left(4r^2 V+f^2\rho^2\right)}.
\end{equation}
Again, the two roots will be equal ($f'_+=f'_-$) if the terms inside the square root equal zero. The resulting equation can be written as a power series of $r$,  
\begin{equation}
\label{36}
2\left[e^2 Q_\rho^2-\left(\epsilon_0\left(f^2-1\right)^2+\epsilon_1\right)\right]+e^2 r^2\left(g^2 Q_f^2-f^2 \rho^2\right)-4e^2 r^4 V=0.
\end{equation}
Eq.~\eqref{36} is satisfied only when
\begin{eqnarray}
V&=&0,\\
Q_\rho&=&\pm\frac{1}{e}\sqrt{ \epsilon_0\left(f^2-1\right)^2+\epsilon_1}=\pm\sqrt{\frac{ \left(f^2-1\right)^2}{g^2}+\frac{\epsilon_1}{e^2}},\\
Q_f&=&\pm\frac{\rho f}{g}.
\end{eqnarray}
Therefore, the BPS equations for this monopole are
\begin{eqnarray}
\rho'&=&\pm \frac{1}{r^2}\sqrt{\frac{ (f^2-1)^2}{g^2}+\frac{\epsilon_1}{e^2}},\label{BPS11}\\
f'&=&\pm \frac{1}{2}g\rho f.\label{BPS12}
\end{eqnarray}

Eqs.~\eqref{BPS11}-\eqref{BPS12} are the main results of our work. They are distinct BPS equations, cannot be reduced to Eqs.~(44) or (50) in the ZZC model~\cite{Zhang:2020xsg}. Note that Eq.~\eqref{BPS11} as well as the energy~\eqref{limitzero} depend on the power of $n$ via $\epsilon_1 = \left(\rho/\rho_0\right)^n$. Choosing different regularization function $\epsilon_1$ amounts to having BPS solutions with smaller energy. In the limit of $n\rightarrow\infty$ the total energy converges to a finite value, $E\rightarrow3.569$ GeV. 

\begin{figure}[h!]
\centering
\includegraphics[width=0.8\linewidth]{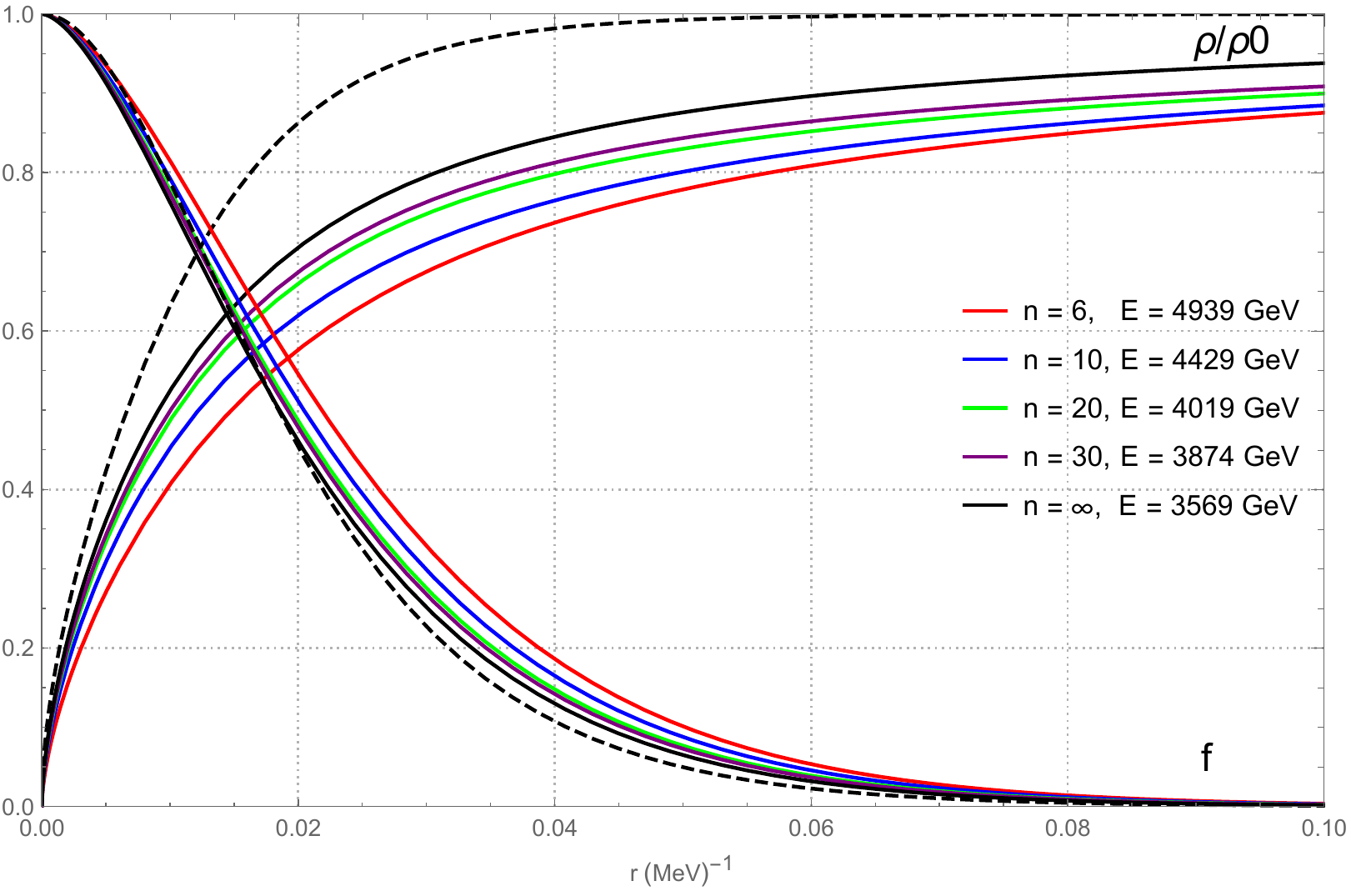}
\caption{The BPS equations \eqref{BPS11}-\eqref{BPS12} for $n = 6, 10, 20, 30$, and $n = \infty$. The solutions for the Eq.~\eqref{eqn:2.12} for $n=6$ (with $A(r)=B(r)=0$) \cite{Zhang:2020xsg} are represented by dashed lines.} 
\label{BPSgraph1combined}
\end{figure}

In Fig.~\ref{BPSgraph1combined} we present the solutions of Eq.~\eqref{BPS11}-\eqref{BPS12} for several values of $n$ up to $n\rightarrow\infty$.
As can be seen, small $n$ produces more diffuse BPS solutions. As $n$ increases the field profiles become more localized. As a comparison, we plot the solution of the $2^{nd}$-order field equations~\eqref{eqn:2.12} for $n=6$. 

The total energy (Eq.~\eqref{eqn:2.14}), after setting $\lambda = 0$, can be written as: 

\begin{eqnarray} \label{energybound}
E&=&4\pi \int_0^\infty dr \left\{\frac{1}{2} r^2 {\rho'}^2 + \frac{1}{2e^2 r^2}\left(\epsilon_0 (f^2-1)^2 +\epsilon_1 \right)+\frac{1}{g^2} {f'}^2+\frac{1}{4} f^2 \rho^2\right\} \nonumber\\
&=&4\pi \int_0^\infty dr \left\{\frac{1}{2} r^2 \left(\frac{1}{r^2}\sqrt{\frac{ (f^2-1)^2}{g^2}+\frac{\epsilon_1}{e^2}}\right)^2 + \frac{\left(\epsilon_0 (f^2-1)^2 +\epsilon_1 \right)}{2e^2 r^2}+\frac{1}{g^2} \left(\frac{1}{2}g\rho f\right)^2+\frac{f^2 \rho^2}{4} \right\}\nonumber\\
&\geq&4\pi \int_0^\infty dr \left\{\frac{ (f^2-1)^2}{g^2 r^2}+\frac{\epsilon_1}{e^2 r^2}
+\frac{f^2 \rho^2}{2} \right\}.
\end{eqnarray}
The bound is saturated by Eqs.~\eqref{BPS11}-\eqref{BPS12}. 

\begin{figure}[h!]
\centering
\includegraphics[width=0.8\linewidth]{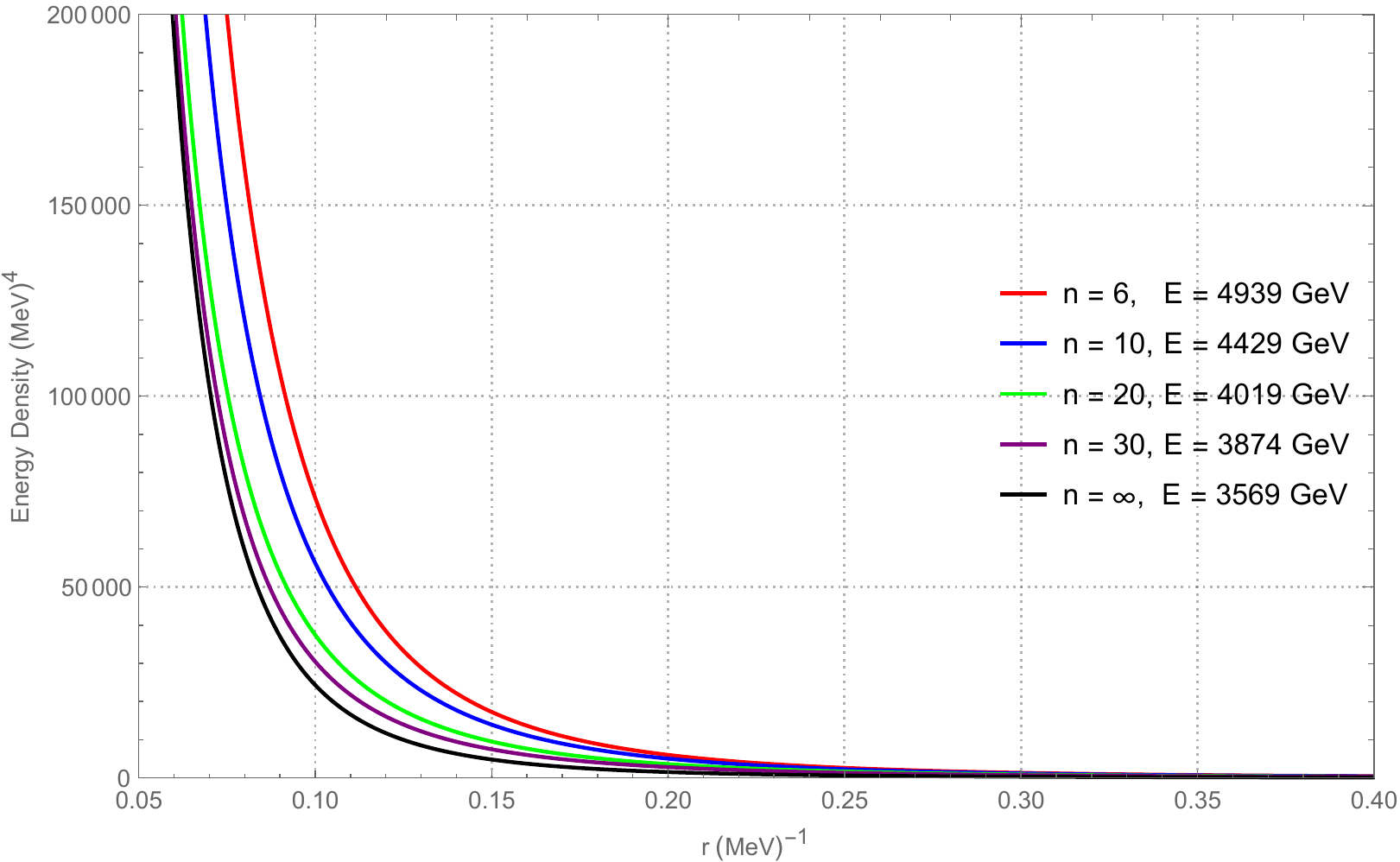}
\caption{The energy density \eqref{energybound} for $n = 6, 10, 20, 30$, and $n = \infty$.} 
\label{energydensitygraph}
\end{figure}

In Fig.~\ref{energydensitygraph} we show the plots of energy densities for several values of $n$. Since $\epsilon_1=(\rho/\rho_0)^n$, we have an energy bound which depends on $n$. The value of this bound decreases as $n$ increases. Note that the energy densities at $r\rightarrow0$ are very high due to the the $\epsilon_1$ term. In the limit where $n\rightarrow\infty$, the integral of $\epsilon_1$ term becomes negligible, and we have the lowest bound for the total energy.

\section{Some Regularization Alternatives for The Cho-Maison Monopole and Their BPS States}
\label{sec:BI}

In the previous section we show how the BPS state of the CM monopole in the ZZC model can be obtained by means of the BPS Lagrangian method. since the ZZC model~\cite{Zhang:2020xsg} is just one toy model to regularize the singularity, it is by no means unique. In this section we shall discuss several other regularization possibilities in the literature and establish their corresponding Bogomolny equations.

\subsection{Born-Infeld Extension in the Hypercharge Sector}

The spirit of ``regularization" seems to lie on adding nonlinear interactions into the field configurations. When talking about nonlinear fields, one natural choice is the Born-Infeld field theory~\cite{Born:1934gh}. Arunasalam and Kobakhidze~\cite{Arunasalam:2017eyu} proposed the the Born-Infeld extension to the hypercharge sector of the Lagrangian ~\eqref{monopolelagr1}:
\begin{eqnarray}
\label{lagrBI1}
\mathcal{L}&=&-|\hat{D}_{\mu}H|^2-\frac{\lambda}{2}\left(H^\dagger H -\frac{\mu^2}{\lambda}\right)^2-\frac{1}{4}F_{\mu\nu}^a F^{a\mu\nu}\nonumber
+\beta^2\left[1-\sqrt{-det\left(\eta_{\mu \nu}+\frac{1}{\beta}G_{\mu \nu}\right)}\right]\nonumber\\
&=&-|\hat{D}_{\mu}H|^2-\frac{\lambda}{2}\left(H^\dagger H -\frac{\mu^2}{\lambda}\right)^2-\frac{1}{4}F_{\mu\nu}^a F^{a\mu\nu}\nonumber\\
&&+\beta^2\left[1-\sqrt{1+\frac{1}{2\beta^2}G_{\mu \nu}G^{\mu \nu}-\frac{1}{16\beta^4}(G_{\mu \nu}\Tilde{G
}^{\mu \nu})^2}\right],
\end{eqnarray}
where $\Tilde{G}^{\mu \nu}\equiv\frac{1}{2}\epsilon^{\mu\nu\alpha\beta}G_{\alpha\beta}$ is the Hodge dual of the field strength tensor $G_{\mu \nu}$, and $\beta$ is the Born-Infeld parameter with the unit of $\textrm{masss}^2$. In the limit $\beta\rightarrow \infty$, the Lagrangian goes back to Eq.~\eqref{monopolelagr1}. 

The field equations are
\begin{eqnarray}
\label{eomBI}
\rho''+\frac{2}{r}\rho'-\frac{f^2}{2r^2}\rho&=&\lambda\left(\frac{\rho^2}{2}-\frac{\mu^2}{\lambda}\right)\rho,\nonumber\\
f''-\frac{f^2-1}{r^2}f&=&\frac{g^2}{4}\rho^2 f,
\end{eqnarray}
which is the same as Eqs.~\eqref{eqn:2.5} when $A=B=0$. The energy is 
\begin{eqnarray}
\label{eqn:2.26}
E&=&E_0 + E_1,\\
E_0&\equiv&\int_0^\infty dr\, \beta^2 \left\{\sqrt{(4\pi r^2)^2+\frac{h_Y^2}{\beta^2}}-4\pi r^2\right\},\\
E_1&\equiv&4\pi \int_0^\infty dr \left\{\frac{(f^2-1)^2}{2g^2 r^2}+\frac{1}{2}(r\rho')^2+\frac{f'^2}{g^2}+\frac{\lambda r^2}{8}(\rho^2-\rho_0^2)^2+\frac{1}{4}f^2\rho^2\right\},
\end{eqnarray}
with $h_Y\equiv4\pi/g'$ the hypermagnetic charge of the monopole. $E_1 \approx 4.1$ TeV as before, while $E_0$ is finite due to the Born-Infeld modification. Its value is
\begin{equation}
\label{EnergiBI}
E_0\approx\frac{\pi^{3/2}}{3\Gamma\left(\frac{3}{4}\right)^2}\sqrt{\frac{\beta h_Y^3}{4\pi}}=\frac{4\pi^{5/2}}{3\Gamma\left(\frac{3}{4}\right)^2}\sqrt{\frac{\beta}{g'^3}}\approx72.8\sqrt{\beta},
\end{equation}
where they use $g'=0.357$. This shows that the monopole energy depends on the parameter $\beta$. The magnetic field is $q_m=4\pi/e$. The numerical value of $\beta$ can be obtained from experiments. The constraint from the Pb-Pb scattering at LHC  \cite{Ellis:2017edi} and considering the $\cos\theta_W$ factor in the sector, $\sqrt{\beta}\gtrsim 90\textrm{ GeV}$. Therefore,
\begin{equation}
E=4.1\textrm{ TeV}+72.8\sqrt{\beta}\approx 10.6\textrm{ TeV}.
\end{equation}
Since the monopole has finite energy, it is then legitimate to ask about its BPS states.

\begin{figure}[h]
\centering
\includegraphics[width=0.8\linewidth]{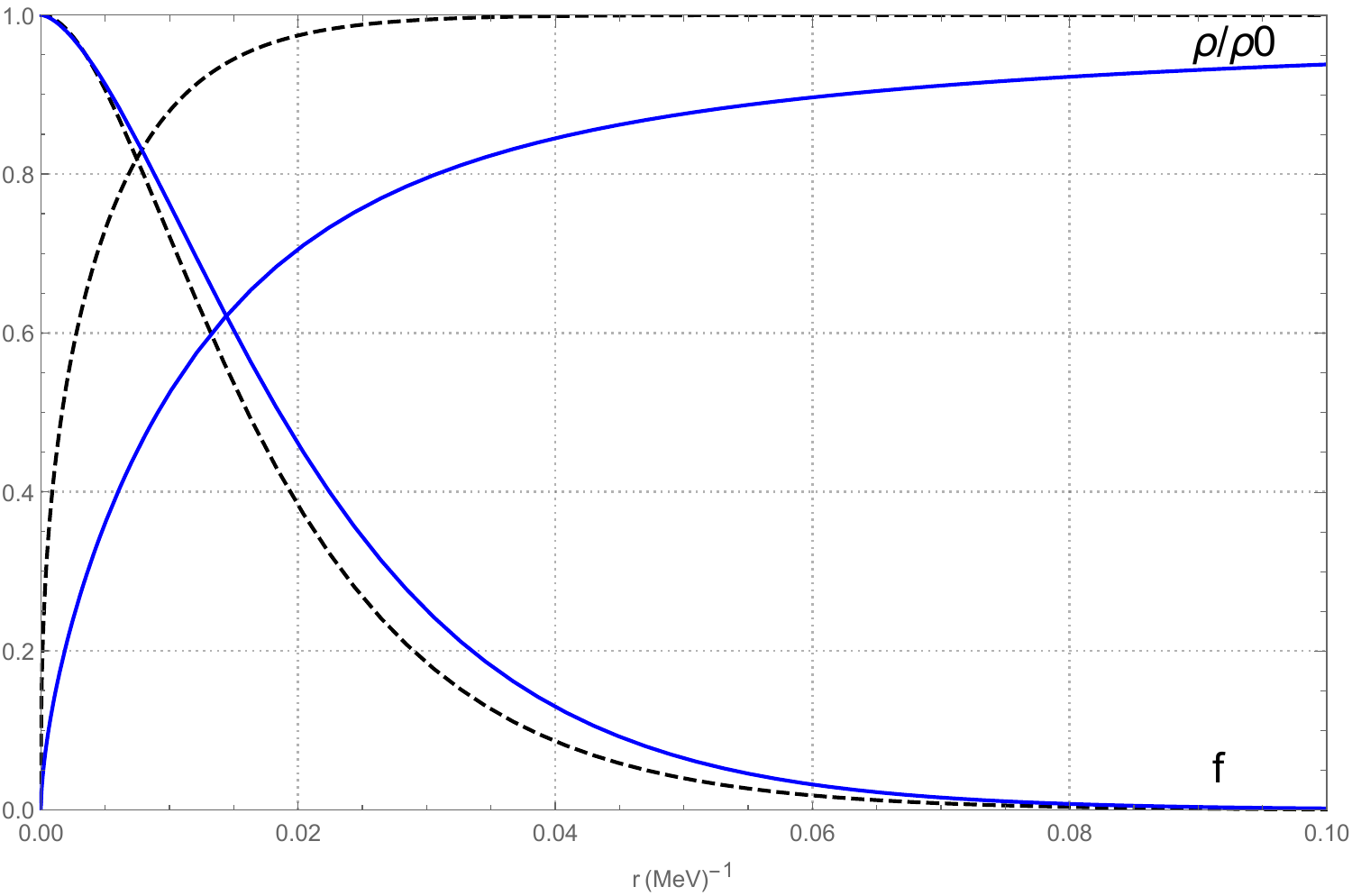}
\caption{\textbf{Solid lines:} The numerical solutions to the BPS equations~\eqref{BPS21}-\eqref{BPS22}.  \textbf{Dashed lines:} The solutions to the Euler-Lagrange equations~\eqref{eomBI}.}
\label{BPSgraph2}
\end{figure}

The Lagrangian (\ref{lagrBI1}) can be written, using the BPS Lagrangian method, as functions of $r$ only:
\begin{eqnarray}
\mathcal{L}&=&\mathcal{L}_s + \mathcal{L}_{BI},\nonumber\\
\mathcal{L}_s&\equiv&-\frac{{\rho'}^2}{2}- \frac{(f^2-1)^2}{2g^2 r^4}-\frac{{f'}^2}{g^2 r^2}-\frac{f^2 \rho^2}{4r^2}-V(\rho),\nonumber\\
\mathcal{L}_{BI}&\equiv&\beta^2 \left(1-\sqrt{1+\frac{1}{{g'}^2 r^4 \beta^2}}\right),
\end{eqnarray}
where $\mathcal{L}_{BI}$ comes from the hypercharge sector. The lowest energy bound is given by~\eqref{EnergiBI}, and  since $\mathcal{L}_{BI}$ is independent of $\rho(r)$ and $f(r)$ it does not enter into the formalism $\mathcal{L}_s-\mathcal{L}_{BPS}=0$. Solving it for $\rho'(r)$, we have
\begin{equation}
\rho'(r)_{\pm}=\frac{Q_\rho}{r^2}\pm \frac{1}{2 r^2}\sqrt{4  Q_\rho^2 - \frac{4}{g^2} (f^2-1)^2-8 r^4 V-2  r^2 f^2 \rho^2 + 8  Q_f r^2 f'-\frac{8r^2 {f'}^2}{g^2}},
\end{equation}
from which the roots of $f'(r)$ can be deduced:
\begin{equation}
\label{persamaanBPS2}
f'(r)_\pm=\frac{g^2 Q_f}{2}\pm \frac{1}{2r}\sqrt{ 2g^2 Q_\rho^2 -2(f^2-1)^2 + g^4 Q_f^2 r^2  - 4g^2 r^4 V -g^2 r^2 \rho^2 f^2}. 
\end{equation}
We then have a polynomial equation in the power of $r$,
\begin{equation}
2\left[g^2 Q_\rho^2 - (f^2-1)^2\right]- 4g^2 r^4 V + g^2 r^2 \left[g^2 Q_f^2 - f^2 \rho^2\right]=0.
\end{equation}
The 4th-order term gives $V=0$. The 0th-order and the 2nd-order terms yield, respectively,
\begin{eqnarray}
Q_\rho&=&\pm \frac{f^2-1}{g},\\
Q_f&=&\pm\frac{\rho f}{g}.
\end{eqnarray}
We have the BPS equations of the form
\begin{eqnarray}
\rho'&=&\pm \frac{f^2-1}{g r^2} \label{BPS21},\\
f'&=&\pm \frac{1}{2} g \rho f \label{BPS22}.
\end{eqnarray}
These results confirm the BPS equations obtained by~\cite{DeFabritiis:2021eah}. Note that the difference between these BPS states and that of t'Hoof-Polyakov's is the appearance of the factor $1/2$, which makes them cannot be solved analytically. We present their numerical solutions in Fig.~\ref{BPSgraph2}. The BPS solutions are rather more diffuse than the corresponding solutions to the Euler-Lagrange field equations~\eqref{eomBI}.    
    

\subsection{Born-Infeld extension in both $SU(2)$ and $U(1)$ sectors}

Modifying the hypercharge sector with nonlinear electrodynamics, despite being useful in terms of the regularization objective, is rather trivial from the BPS point of view. This is because such term does not enter in the BPS condition~\eqref{l-l}. A more interesting case is when the Born-Infeld extension is not applied only to the hypercharge but also to the $SU(2)$ Yang-Mills sectors. This model was proposed by Arunasalam, Collison, and Kobakhidze (ACK)~\cite{Arunasalam:2018iom}.

The corresponding Lagrangian can be written as
\begin{eqnarray}
\label{ACK}
\mathcal{L}&=&\mathcal{L}_s + \mathcal{L}_{BI1}+\mathcal{L}_{BI2},\nonumber\\
\mathcal{L}_s&\equiv&-\frac{{\rho'}^2}{2}-\frac{f^2 \rho^2}{4r^2}-V(\rho),\nonumber\\
\mathcal{L}_{BI1}&\equiv&\beta_1^2 \left(1-\sqrt{1+\frac{(f^2-1)^2}{g^2 \beta_1^2 r^4}+\frac{2{f'}^2}{g^2 \beta_1^2 r^2}}\right),\nonumber\\
\mathcal{L}_{BI2}&\equiv&\beta_2^2 \left(1-\sqrt{1+\frac{1}{{g'}^2 r^4 \beta_2^2}}\right).
\end{eqnarray}
Since, like before, $\mathcal{L}_{BI2}$ does not contribute to the field solutions, the BPS equations can be written as $(\mathcal{L}_s +\mathcal{L}_{BI1})-\mathcal{L}_{BPS}=0$. Solving it for $\rho'(r)$,
\begin{equation}
\rho'_{\pm} =\frac{Q_\rho}{r^2}\pm \sqrt{\frac{Q_\rho^2}{r^4}+2\left(V - \beta_1^2 + \frac{f^2 \rho^2}{4r^2}-\frac{Q_f f'}{r^2}+\sqrt{\beta_1^4+\frac{1}{2}\beta_1^2 \left(\frac{2(f^2-1)^2}{g^2 r^4}+\frac{4 {f'}^2}{g^2 r^2}\right)}\right)}.
\end{equation}
This can be used to solve $f'(r)$,
\begin{equation}
f'_\pm = \frac{K+\sqrt{\frac{128 \beta_1^2}{r^2} D_2}}{L},
\end{equation}
where
\begin{eqnarray}
D_2 &\equiv& 4 g^2 Q_\rho^4 - 16 g^2 r^8 (2 \beta_1^2 - V) V +4 g^2 r^2 (2 Q_f^2 - 4 Q_f^2 f^2 + 2 Q_f^2 f^4 - Q_\rho^2 f^2 \rho^2)\nonumber\\
&&+ r^4 (-16 \beta_1^2 + 16 g^2 Q_\rho^2 \beta_1^2 + 32 \beta_1^2 f^2 - 16 \beta_1^2 f^4 - 16 g^2 Q_\rho^2 V + g^2 f^4 \rho^4)\nonumber\\
&&+ 8 g^2 r^6 (g^2 Q_f^2 \beta_1^2 - \beta_1^2 f^2 \rho^2 + f^2 V \rho^2),\nonumber\\
K&\equiv& - g^2 Q_f \left(\frac{16 Q_\rho^2}{r^2} + 32 r^2 \beta_1^2 - 32 r^2 V - 8 f^2 \rho^2 \right),\nonumber\\
L&\equiv& 2 (16 g^2 Q_f^2 - 32 r^2 \beta_1^2).
\end{eqnarray}

Setting $D_2=0$, the cofficient for the 8th-order power constraints the potential to be constant, $V=0$ or $2 \beta_1^2$. From the 6th-order term we have 
\begin{equation}
Q_f=\pm \frac{\rho f \sqrt{\beta_1^2 - V}}{g \beta_1},
\end{equation}
Choosing $V=2 \beta_1^2$ gives imaginary value for $Q_f$, so we choose $V=0$. Substituting these results to $D_2$ gives
\begin{eqnarray}
\label{DD}
D_2&=& 4 g^2 Q_\rho^4+4 g^2 r^2 (2 Q_f^2 - 4 Q_f^2 f^2 + 2 Q_f^2 f^4 - Q_\rho^2 f^2 \rho^2) \nonumber\\
&&+r^4 (-16 \beta_1^2 + 16 g^2 Q_\rho^2 \beta_1^2 + 32 \beta_1^2 f^2 - 16 \beta_1^2 f^4 + g^2 f^4 \rho^4),
\end{eqnarray}
with
\begin{equation}
Q_f=\pm \frac{\rho f}{g}.\nonumber
\end{equation}
We can immediately see that there is an incompatibility with nontrivial BPS equations here. The zeroth-order power of Eq.~\eqref{DD} implies $Q_\rho=0$, while the 2nd- and the 4th-order terms yield, respectively, 
\begin{eqnarray}
Q_\rho &=&\pm\left\{\frac{\sqrt{2} (f^2-1)}{g},\ \frac{\sqrt{16 \beta_1^2 (f^2-1)^2 - g^2 \rho^2 f^2}}{4 g \beta_1}\right\}.
\end{eqnarray}
These three $Q_{\rho}$ conditions cannot be consistently satisfied unless $f(r)$ and $\rho(r)$ are constants, 
\begin{equation}
\label{trivial}
f=\pm1,\ \ \ \rho=0.
\end{equation}
We conclude that within this BPS Lagrangian method, there is no BPS states for the ACK model of electroweak monopole.


\subsection{Generalized electroweak monopole}

As our last attempt, we can generalize the ``regulators" in Eq.~\eqref{eqn:2.11} by introducing two functions dependent to the scalar field, $W(\rho)$ and $G(\rho)$. These act as the multiplier to the Lagrangian, which will generalize the model from equation \eqref{eqn:2.11}. In the context of t'Hooft-Polyakov monopole this mechanism has been discussed by authors in~\cite{Casana:2012un, Casana:2013lna}, and its BPS Lagrangian method was elaborated in~\cite{Atmaja:2018cod}.

The Lagrangian, in the case of monopole ($A(r)=B(r)=0$), is
\begin{eqnarray}
\label{generallagr}
\mathcal{L}&=&-G(\rho)\left(\frac{1}{2}(\partial_\mu \rho)^2+\frac{g^2}{4}\rho^2 W_\mu^* W^\mu  \right)
-V(\rho)-W(\rho) \bigg(\frac{1}{4}\epsilon(\rho){F_{\mu\nu}^{(em)}}^2\nonumber\\
&&+\frac{1}{2}\bigg|D_{\mu}^{(em)}W_\nu-D_{\nu}^{(em)} W_\mu \bigg|^2
-ie F_{\mu \nu}^{(em)}{W^*}^\mu W^\nu -\frac{g^2}{4}(W^*_\mu W_\nu-W^*_\nu W_\mu)^2\bigg).
\end{eqnarray}
This is the generalized version of the nonsingular electroweak monopole with non-vacuum electromagnetic permittivity \eqref{eqn:2.11}. 
Using the ansatz~\eqref{ansatzem}, the Lagrangian reads
\begin{equation}
\label{lagrgeneral1}
\mathcal{L}=-G(\rho) \left(\frac{{\rho'}^2}{2} +\frac{f^2 \rho^2}{4r^2}\right)
- W(\rho)\left[ \frac{1}{2e^2 r^4}\left\{\epsilon_0 (f^2-1)^2 +\epsilon_1 \right\}+\frac{{f'}^2}{g^2 r^2}\right]-V
\end{equation}
From $\mathcal{L}-\mathcal{L}_{BPS}=0$ we have 
\begin{eqnarray}
\rho'_\pm&=&\frac{Q_\rho}{G r^2}\pm\frac{1}{2 G r^2}\bigg\{4 Q_\rho^2-G \bigg(8r^4 V+\frac{4}{e^2}  W\left(\epsilon_0(f^2-1)^2+\epsilon_1\right)\nonumber\\
&&+2 G r^2 f^2 \rho^2 -8 Q_f r^2 f'+\frac{8 r^2 W {f'}^2}{g^2}\bigg)\bigg\}^\frac{1}{2}.
\end{eqnarray}
Solving for $f'(r)$, we have
\begin{eqnarray}
f'_\pm&=&\frac{g^2 Q_f}{2 W}\pm \frac{g}{e \sqrt{G} r W}\bigg\{e^2\left(2 W Q_\rho^2 +g^2 G Q_f^2 r^2\right)\nonumber\\
&&-2GW^2\left(\epsilon_0 \left(f^2-1\right)^2+\epsilon_1\right)-e^2 r^2 G  W\left(4r^2 V +f^2 G \rho^2\right)\bigg\}^\frac{1}{2}.
\end{eqnarray}
The terms inside the square root satisfies
\begin{eqnarray}
-32 e^4 G r^8 V W^2 + 
 16 e^2 r^4 W^2 \bigg(e^2 Q_\rho^2 - G W \left(\epsilon_0 (f^2-1)^2 + \epsilon_1\right)\bigg)
 \nonumber\\ 
  + 8 e^4 G r^6 W\left(g^2 Q_f^2 - f^2 G W \rho^2\right)=0.
\end{eqnarray}
The 8th-order term yields $V=0$. The 4th-order and the 6th-order term give, respectively,
\begin{eqnarray}
Q_\rho&=&\pm \sqrt{G W}\sqrt{\frac{ (f^2-1)^2}{g^2}+\frac{\epsilon_1}{e^2}},\\
Q_f&=&\pm\frac{\sqrt{G W}\rho f}{g}.
\end{eqnarray}
The BPS equations are thus
\begin{eqnarray}
\label{WGbps1}
\rho'&=&\pm \frac{1}{r^2} \sqrt{\frac{W}{G}} \sqrt{\frac{ (f^2-1)^2}{g^2}+\frac{\epsilon_1}{e^2}},\\
\label{WGbps2}
f'&=&\pm \frac{1}{2}g\rho f \sqrt{\frac{G}{W}}.
\end{eqnarray}

The energy bound of this monopole can be found in a similar way as in Eq.~\eqref{energybound}. Substituting Eqs.~\eqref{WGbps1}\eqref{WGbps2} to the Lagrangian~\eqref{generallagr} and integrating it gives
\begin{eqnarray}
E&=& 4\pi \int_0^\infty dr\left\{ 
G(\rho) \left[\frac{r^2{\rho'}^2}{2} +\frac{f^2 \rho^2}{4}\right]
+ W(\rho)\left[ \frac{1}{2e^2 r^2}\left\{\epsilon_0 (f^2-1)^2 +\epsilon_1 \right\}+\frac{{f'}^2}{g^2 }\right]
\right\}\nonumber\\
&=& 4\pi \int_0^\infty dr 
\left\{G(\rho) \left[\frac{W(\rho)}{2 G(\rho) r^2}\left(\frac{ (f^2-1)^2}{g^2}+\frac{\epsilon_1}{e^2}\right)
+\frac{f^2 \rho^2}{4}\right] \right.
\nonumber \\
& & \qquad \qquad \quad \left. 
+ W(\rho)\left[ \frac{1}{2e^2 r^2}\left\{\epsilon_0 (f^2-1)^2 +\epsilon_1 \right\}+\frac{G(\rho)}{W(\rho)} \frac{\rho^2 f^2}{4} \right]
\right\}\nonumber\\
&\geq&
4\pi \int_0^\infty dr
\left\{W(\rho)
\left(
\frac{ (f^2-1)^2}{g^2 r^2}+\frac{\epsilon_1}{e^2 r^2}
\right)
+G(\rho) \frac{f^2 \rho^2}{2}
\right\}.
\end{eqnarray}
The bound is determined by the forms of $G(\rho)$ and $W(\rho)$.



\section{Conclusions}
\label{sec:conc}

The possible existence of magnetic monopole in electroweak theory is what makes the Cho-Maison monopole appealing theoretically. However, its singular energy problem must be resolved before any serious experimental result can be claimed. Zhang, Zhou, and Cho in~\cite{Zhang:2020xsg} regularize the electroweak monopole by means of renormalization of the ``bare" electromagnetic permittivity. Once the finite energy is achieved it is just natural to ask whether the BPS solutions that saturate the lowest energy bound exist. In this paper we employ the BPS Lagrangian method to systematically generate such solutions.

The nature of electroweak Cho-Maison monopole opens up degeneracy in its BPS states, not found in the t'Hooft-Polyakov case. ZZC reported that for a given regularized Lagrangian there can exist at least two sets of BPS equations, each having different energy bound. Using the BPS Lagrangian method, we report a new set of BPS equations within the ZZC model, Eqs.~\eqref{BPS11}-\eqref{BPS12}. These equations are not reducible to the known BPS states in ZZC paper~\cite{Zhang:2020xsg}. The solutions depends on $n$, the polynomial power of the permittivity expansion. The corresponding energy bound (in the limit of $n\rightarrow\infty$) is $E\simeq3.569$ GeV. This value is higher than the lowest energy bound predicted by ZZC, but still within the allowed theoretical window.

In this work we also consider several other regularization proposals and apply the BPS method to extract their corresponding BPS states. The first is the Born-Infeld extension in the hypercharge sector~\cite{Arunasalam:2017eyu}. In this model the Born-Infeld modification does not enter the field equations but does regularize the energy. We obtain the BPS equations \eqref{BPS21}-\eqref{BPS22}, which is the same as obtained in~\cite{DeFabritiis:2021eah}. They are just different from the t'Hooft-Polyakov ones by a factor of $1/2$. An immediate extension to the first model is an electroweak monopole with Born-Infeld form in both the hypercharge and Yang-Mills sectors. This model has been considered in~\cite{Arunasalam:2018iom}. The $SO(3)$ counterpart of this monopole can be found, for example,  in~\cite{Grandi:1999rv}. Our investigation reveals that within the BPS Lagrangian method no nontrivial configuration for $\rho(r)$ and $f(r)$ exist. It seems that having Born-Infeld form in both the gauge sectors is too restricted for the existence of the BPS state. For the last model we study, we construct a ``{\it generalized} electroweak monopole" in the same spirit as in~\cite{Casana:2012un, Casana:2013lna}, having non-canonical kinetic terms whose generalized functions $G$ and $W$ depend only on the Higgs field and not on its derivative. The resulting BPS equations have the factor of $\sqrt{W/G}$ in both $\rho'(r)$ and $1/f'(r)$, as shown in Eqs.~\eqref{WGbps1}-\eqref{WGbps2}.

Finally we should comment on the limitation of the Bogomolny method we employ. It is surprising that the method gives us a new set of distinct BPS equations, Eqs.~\eqref{BPS11}-\eqref{BPS12}. But what is more intriguing is that they are the only BPS solutions for the ZZC model of electroweak monopole using this mechanism. Surely this means that the method cannot probe the possible existence of other BPS solutions. Secondly, it is well-known that every topological soliton solutions is labeled by their topological charge, and that there is a relation between the BPS energy and the charge. However, in this BPS method the direct relation seems to be absent. The BPS Lagrangian $\mathcal{L}_{BPS}$ cannot be expressed as a total differential, whose integral does not give us the boundary term ({\it i.e,} the topological charge),
\begin{equation}
E_{BPS}\neq Q.
\end{equation}
Lastly, this method fails to give us the BPS equations for the ACK model~\cite{Arunasalam:2018iom} described by the Lagrangian~\eqref{ACK}. The method leass to three incompatible conditions for the function $Q_{\rho}$, which amount to having trivial solutions~\eqref{trivial}. At the moment it is not yet clear whether this failure is due to the BPS method or the structure of the non-Abelian Born-Infeld monopole itself. The Grandi-Moreno-Schaposnik monopole~\cite{Grandi:1999rv} is not known to have BPS states. In any case, Atmaja in~\cite{Atmaja:2018cod} gives suggestion on how to generalize the $\mathcal{L}_{BPS}$: by introducing {\it non-boundary terms}. These terms can later be determined from the constraints of the system. It would be interesting to see whether this approach can be applied to the electroweak monopole case to cure our problems. This wil be the topic of our forthcoming publications.

\begin{acknowledgments}
AG and HSR are supported by Universitas Indonesia's Hibah Riset MIPA UI No.~PKS-026/UN2.F3.D/PPM.00.02.2023, while IP is funded by Sampoerna University's Internal Research Grant No. 003/IRG/SU/AY.2023-2024.
\end{acknowledgments}


\end{document}